\documentclass{aa}  

\usepackage{graphicx}
\usepackage{txfonts}
\usepackage{amsmath}
\usepackage{amssymb}
\usepackage{color}
\usepackage{natbib}
\usepackage{hyperref}
\makeatletter
\renewcommand*\aa@pageof{, page \thepage{} of \pageref*{LastPage}}
\makeatother

\newcommand{\Msun}{$\mathrm{M}_{\odot}$}
\newcommand{\Rsun}{$\mathrm{R}_{\odot}$}
\newcommand{\Lsun}{$\mathrm{L}_{\odot}$}

\newcommand*\mean[1]{\overline{#1}}

\newcommand{\mlt}{SSM-MLT}
\newcommand{\eq}{SSM-K1}
\newcommand{\dcc}{$\delta c^2/c^2$}

\defcitealias{Magg2022}{M22}
\defcitealias{Asplund2009}{A09}
\defcitealias{Grevesse1998}{GS98}

\begin{document}

   \title{Testing a non-local 1-equation turbulent convection model:\\A solar model}

%   \subtitle{subtitle}

   \author{T. A. M. Braun
          \inst{1, 2}\thanks{braun@mpa-garching.mpg.de}
          \and
          F. Ahlborn\inst{3}
        \and
        A. Weiss\inst{1, 2}
          }

   \institute{Max-Planck-Institut f{\"u}r Astrophysik, Karl-Schwarzschild-Straße 1, 85741 Garching, Germany
    \and
    Ludwig-Maximillians-Universit\"at M\"unchen, Geschwister-Scholl-Platz 1, 80539 Munich, Germany
    \and
    Heidelberger Institut für Theoretische Studien, Schloss-Wolfsbrunnenweg 35, 69118 Heidelberg, Germany\\
             }

   \date{}

% \abstract{}{}{}{}{} 
% 5 {} token are mandatory
 
  \abstract
   {Turbulent convection models treat stellar convection more physically than standard mixing-length theory by including non-local effects. We recently successfully applied the Kuhfuss version to convective cores in main sequence stars. Its usefulness for convective envelopes remains to be tested.}
   {The solar convective envelope constitutes a viable test bed for investigating the usefulness of the 1-equation Kuhfuss turbulent convection model.}
   {We used the one-dimensional stellar evolution code GARSTEC to calculate a standard solar model with the 1-equation Kuhfuss turbulent convection model, and compared it to helioseismic measurements and a solar model using standard mixing-length theory. Additionally, we investigated the influence of the additional free parameters of the convection model on the solar structure.}
   {The 1-equation Kuhfuss model reproduces the sound-speed profile and the lower boundary of the convective region less well than the mixing-length model, because the inherent non-local effects overestimate the amount of convective penetration below the Schwarzschild boundary. We trace this back to the coupling of the temperature gradient to the convective flux in the 1-equation version of the Kuhfuss theory.}
   {The temperature stratification of the solar convective envelope is not well modelled by the 1-equation Kuhfuss turbulent convection model, and the more complex 3-equation version is needed to improve the modelling of convection in the envelopes of 1D stellar evolution models.}


   \maketitle
%
%-------------------------------------------------------------------

%%%%%%%%%%%%%%%%%%%%%%%%%%%%%%%%%%%%%%%%%%%%%%%%%%%%%%%%%%%%
\section{Introduction}\label{sec:intro}
%%%%%%%%%%%%%%%%%%%%%%%%%%%%%%%%%%%%%%%%%%%%%%%%%%%%%%%%%%%%

Convection plays a major role in the evolution of a star, as it is one of the main transport mechanisms for energy and chemical elements. However, modelling convection remains a challenge, because convection is an inherently three-dimensional and highly turbulent process. 

%3D
Three-dimensional (3D) hydrodynamic simulations of wedges or boxes in stars over a limited time span provide an opportunity to study the features of convection \citep[e.g. ][]{Chiavassa2011, Trampedach2014, Magic2015, Kapyla2017, Andrassy2024, Herwig2023}. However, simulation of the entire evolution of a star remains an impossible task.
Even with the expected increase in computing power in the near future, it will not be possible to follow the entire evolution of a star in a 3D simulation, as the simulation needs to cover many orders of magnitude in length scale and timescale.
Simulating a thermal timescale is already a difficult task, and the simulation result may still depend on the initial thermal structure.

%1D
One-dimensional (1D) stellar-evolution
codes still provide the only possibility to follow the long-term evolution of stars. While the numerical methods and treatment of physics have constantly been refined, from the first attempts to model this evolution \citep{Schwarzschild1953,Schwarzschild1957,Hofmeister1964,Iben1965} to the more recent efforts \citep[e.g. ][]{Dotter2008, Weiss2008, Jermyn2023}, the method of choice to deal with convection has been, and still is, mixing length theory (MLT), which dates back to \citet{Prandtl1925} and \citet{Biermann1932} in terms of the physical picture, and is predominantly used in the versions by \citet{Bohm-Vitense1958} and \citet{Cox1968}.

%MLT
Mixing
length theory uses the approximation of a bubble of hot material that rises and eventually dissolves in cooler regions, thus transporting energy and chemical elements. 
Due to its adjustable parameter ---the mixing length---, which describes the average distance travelled by a bubble, it successfully reproduces the overall structure of convective layers both in the cores of massive stars and in the envelopes of cool stars. 
However, its simplicity neglects key features of convection, namely non-locality and time dependence. 
As the adjustment of convection to changes in the stellar structure happens on a dynamical timescale, the approximation of instantaneous adjustment (and mixing of the chemical species) is reasonable as long as the structural changes happen on a longer timescale, such as the thermal or even more so the nuclear timescale during core hydrogen and helium burning.
This approximation becomes questionable for events with timescales shorter than or comparable to the timescale of convection, such as the core He-flash in low-mass stars \citep{Flaskamp2003}. Furthermore, the time dependence is important for stellar pulsations \citep[e.g. ][]{Feuchtinger1998, Smolec2008}.

%CBM
A more crucial weakness of MLT is the inherent assumption of locality. 
At the boundary of convectively unstable regions, that is at the radius where the acceleration due to the radial buoyancy force drops to zero (the so-called Schwarzschild or Ledoux boundary), MLT predicts zero convective velocity by construction. However, the material approaches this boundary with non-zero velocity, and will penetrate into layers that are locally stable against convection. This convective boundary mixing (CBM) can potentially modify both the chemical composition, and the temperature gradient in the regions adjacent to the convectively unstable region \citep[see the review by ][]{Anders2023}. 
There is a long list of cases where CBM provides a way to reconcile stellar models with observations.
Examples are the width of the main sequence (MS) \citep{Napiwotzki1991}, the morphology of open-cluster colour--magnitude diagrams \citep{Maeder1981, Demarque1994}, or the requirement of identical ages for the components of detached eclipsing binaries \citep{Claret2016}. 
Furthermore, the overestimation of the luminosity of the red giant bump in globular clusters by stellar models using standard MLT \citep{King1985, Alongi1991} and the `Cephe{\"\i}d mass discrepancy problem' \citep{Stobie1969} can be explained by CBM.
Here, the mass estimate based on the pulsation and the mass derived from evolutionary models that match the observed luminosity disagree. The pulsational masses have recently been shown to agree with the dynamical masses in LMC Cephe{\"\i}d binaries \citep{Pietrzynski2010}. Additional mixing beyond the stability limit into the CBM region is needed to solve this mass discrepancy \citep{Chiosi1992, Cassisi2011, Neilson2011}.

%overshooting
To solve these discrepancies between observations and stellar models, 1D stellar evolution codes commonly incorporate CBM as an additional parametrised overshooting of convection. This adds the mixing of chemical elements without heat transport, which means the temperature gradient is not modified in the CBM region. 
This is either described as complete mixing up to a certain fraction of the scale height, or as a diffusive process \citep{Freytag1996}, with an exponentially decaying mixing speed and again scaled by an appropriate scale height.
In the following, we refer to this treatment of CBM as `ad hoc overshooting'.

%Kuhfuss model
A more physical but still one-dimensional description of CBM is provided by non-local turbulent convection models (TCMs). In contrast to ad hoc overshooting, TCMs are derived from the fundamental hydrodynamic equations and allow us to predict the structure and extent of the CBM region.
Several non-local TCMs have been developed \citep[e.g. ][]{Xiong1997, Canuto1998, Li2007}, with the Kuhfuss model \citep{Kuhfuss1986, Kuhfuss1987}, which is tested in this work, being one of them. 
There are two versions of the Kuhfuss model: the 3-equation model and the simplified 1-equation model (see Sect.~\ref{sec:kuhfuss} for more details). 
Both versions were improved \citep{Wuchterl1998, Flaskamp2003, Kupka2022, Ahlborn2022}, and implemented in the GARching STellar Evolution Code \citep[][GARSTEC]{Weiss2008}. The results for MS stars in the mass range of 1.5 to 8~\Msun\ with convective cores are in qualitative agreement with models using MLT and ad hoc overshooting, which were tuned to match observations \citep{Ahlborn2022}. 
In particular, we showed that the extent of the chemically mixed core, as obtained from using the 1-equation version of the Kuhfuss model, is in very good agreement with that resulting from the 3-equation version, while the two versions lead to different temperature gradients \citep{Kupka2022}. This encouraging first result led us to consider further tests of the convection model for stars of different mass and different evolutionary stages, including the Cephe{\"\i}d mass discrepancy (Deka et al., in preparation).
%sun
A first such test case for the Kuhfuss model is convection in stellar envelopes, such as those in lower mass MS stars like the Sun. 
The conditions under which convection in stellar envelopes happens are different from the conditions in cores. It is assumed that convection in envelopes is driven by cooling from the surface \citep{Spruit1997}, while convection in cores is driven by the heating from the hottest layers in the centre of the star. The temperature stratification in convective cores is nearly adiabatic, while convective envelopes can be extremely superadiabatic, the temperature gradient sometimes being nearer to the radiative than to the adiabatic value.
Furthermore, the density contrast between the top and the bottom of the convective region is larger in convective envelopes compared to convective cores. Convective envelopes consist of narrow, fast downdrafts driven by cooling, and upflows driven by mass conservation, which replace the mass that flows down \citep{Stein1989, Trampedach2010, Trampedach2014}. 
Therefore, applying and testing the Kuhfuss model with convective envelopes will allow us to gauge the extent of its applicability.

The solar envelope is the prime test case for stellar models, because its structure ---including the depth of the convective region--- inferred from helioseismology \citep[see][and Sect.~\ref{sec:sun-observation}]{Basu2016} is well known. In this paper, we therefore use the Sun to study the predictions of the 1-equation model for the solar convective envelope and its structure.
After a short introduction to the Kuhfuss 1-equation model in Sect.~\ref{sec:kuhfuss}, the solar models and available observable quantities are described in Sect.~\ref{sec:sun}. We present our results in Sect.~\ref{sec:results}, and provide a discussion of these results in Sect.~\ref{sec:discussion}. We end the paper with our conclusions and a summary (Sect.~\ref{sec:conclusion+summary}).

%%%%%%%%%%%%%%%%%%%%%%%%%%%%%%%%%%%%%%%%%%%%%%%%%%%%%%%%%%%%
\section{The Kuhfuss 1-equation model}\label{sec:kuhfuss}
%%%%%%%%%%%%%%%%%%%%%%%%%%%%%%%%%%%%%%%%%%%%%%%%%%%%%%%%%%%%

\citet{Kuhfuss1986} derived a turbulent convection model based on the Reynolds averaged hydrodynamic conservation equations \citep{Reynolds1895}.
To compute the Reynolds averaged equations, the quantities of a fluid are split into a fluctuating part $a'$ and a spherically averaged part $\mean{a}$. This procedure is also called Reynolds splitting.
\citet{Kuhfuss1986} applied this splitting to the quantities of the fluid and derived dynamical equations for the turbulent kinetic energy (TKE) $\omega=\frac{1}{2}\mean{\boldsymbol{u}'^2}$, the convective flux variable $\Pi=\mean{s'u'_r}$, and the second-order entropy fluctuations $\Phi=\frac{1}{2}\mean{s'^2}$. The variables $s'$ and $\boldsymbol{u}'$ denote the fluctuations of specific entropy and velocity. The radial component of $\boldsymbol{u}'$ is denoted $u'_r$. From here on, the mean values are given without bars for improved readability.

The relation between $\Pi$ and the convective flux is given by $F_\mathrm{conv} = \rho T \Pi$, where $\rho$ and $T$ denote mean density and temperature.
The dynamical equations for $\omega$, $\Pi$, and $\Phi$ constitute the 3-equation model. 
In this work, the more simplified 1-equation model is used, which is summarised below. For more details on the one- and 3-equation models, we refer the reader to the original work of \citet{Kuhfuss1987}, and the improvements introduced by \citet{Kupka2022} and \citet{Ahlborn2022}.

The main difference between the one- and the 3-equation model is the downgradient approximation, which is applied to the convective flux in the 1-equation model. In analogy to Fick's law of diffusion, the convective flux is assumed to be proportional to the gradient of the entropy. This approximation is also applied in MLT. In the case of the 1-equation model,  for the convective flux, this approximation implies: 
\begin{equation}\label{eq:downgradientapprox_pi}
    \Pi = -\alpha_\mathrm{s} \Lambda \omega^{1/2} \frac{\partial s}{\partial r} \, ,
\end{equation}
with a free parameter $\alpha_\mathrm{s}$, and a characteristic length scale $\Lambda$.
The gradient of the specific entropy can be described by
\begin{equation}\label{eq:entropy_gradient}
    \frac{\partial s}{\partial r} = - \frac{c_\mathrm{p}}{H_\mathrm{p}}(\nabla - \nabla_\mathrm{ad}) \, ,
\end{equation}
with the model and the adiabatic temperature gradient, $\nabla$ and $\nabla_\mathrm{ad}$, respectively, the specific heat $c_\mathrm{p}$, and the pressure scale height $H_\mathrm{p}$.
This reduces the model to one equation, namely to the equation describing the TKE, $\omega$:
\begin{equation}\label{eq:1eq}
    \frac{\partial \omega}{\partial t} = \frac{\nabla_\mathrm{ad}T\Lambda \alpha_\mathrm{s} c_\mathrm{p}}{H_\mathrm{p}^2}\sqrt{\omega}(\nabla-\nabla_\mathrm{ad}) - \frac{C_\mathrm{D}}{\Lambda}\omega^{3/2}-\frac{\omega}{\tau_\mathrm{rad}}-\mathcal{F}_\omega \, .
\end{equation}
The second and third terms in this equation describe the viscous dissipation with the free parameter $C_\mathrm{D}$ and the radiative dissipation, with
\begin{equation}\label{eq:tau}
    \tau_\mathrm{rad} = \frac{c_\mathrm{p} \kappa \rho^2 \Lambda^2}{4 \sigma T^3 \gamma^2_\mathrm{R}} \, ,
\end{equation}
with the opacity $\kappa$, the Stefan-Boltzmann constant $\sigma$, and the free parameter $\gamma_\mathrm{R}$. The inclusion of the radiative losses in the 1-equation model is an extension of the model by \citet{Wuchterl1998}.

The fourth term of Eq.~(\ref{eq:1eq}) describes the non-local transport processes, which are modelled with a downgradient approximation
\begin{equation}\label{eq:non-local_effects}
    \mathcal{F}_\omega = -\frac{1}{\rho}\boldsymbol{\nabla} \cdot (\alpha_\omega \rho \Lambda \sqrt{\omega} \boldsymbol{\nabla} \omega) \, ,
\end{equation}
introducing another free parameter $\alpha_\omega$.

Given the equation for the convective flux (Eq.~\ref{eq:downgradientapprox_pi} and \ref{eq:entropy_gradient}), the temperature gradient for the model can be calculated as
\begin{equation}\label{eq:tempgradient}
    \nabla - \nabla_\mathrm{ad} = {(\nabla_\mathrm{rad} - \nabla_\mathrm{ad})}\left({1 + \frac{\rho c_\mathrm{p} \alpha_\mathrm{s} \Lambda \sqrt{\omega}}{k_\mathrm{rad}}}\right)^{-1} \, ,
\end{equation}
where $\nabla_\mathrm{rad}$ denotes the temperature gradient that is expected if all energy is transported by radiation and $k_{\rm rad}$ denotes the radiative conduction coefficient \citep[see e.g.][Eq. 5.10]{Kippenhahn2013}. 

In analogy to MLT, the characteristic length scale $\Lambda$ can be expressed by a parameter $\alpha_\Lambda$ and the pressure scale height $H_\mathrm{p}$. \citet{Wuchterl1995} introduced another parameter $\beta$ to account for cases where $H_\mathrm{p}$ diverges, as is the case for example towards the centre, so that $\Lambda$ is modified according to

\begin{equation}\label{eq:lambda_limiting}
    \frac{1}{\Lambda} = \frac{1}{\alpha_\Lambda H_\mathrm{p}} + \frac{1}{\beta r} \, .
\end{equation}

Compared to MLT, the Kuhfuss 1-equation model has the advantage of having both the non-locality and the time dependency of convection already included in the theory. As in \citet{Kupka2022} and \citet{Ahlborn2022}, we are interested in the stationary situation only, setting the time derivative on the left-hand side of Eq.~\ref{eq:1eq} to zero. A time-dependent but local version of the 1-equation model to deal with stellar pulsations was implemented in the MESA-code by \citet{mesarsp:2019}.

In the non-local version, as implemented in GARSTEC, the CBM emerges naturally from the solution of the model equations \citep{Ahlborn2022}, without applying ad hoc overshooting, as needed in MLT.
An important difference between the two convection models is that the free parameters of the 1-equation Kuhfuss model are explicitly stated and are connected to distinct physical processes. 
While several numerical parameters are introduced in the derivation of MLT, they are generally not modified when MLT is applied. In most applications of MLT, there is only one free parameter left: the mixing length parameter, which controls the efficiency of convective transport. In contrast, our implementation of the 1-equation model allows us to vary and test the impact of all free parameters of the theory. Even though this also poses the challenge of calibrating a larger number of parameters, we consider this an advantage of the implementation of the 1-equation model.

\begin{table}[htb]
\caption{Default parameters of the Kuhfuss 1-equation model}  
\label{tab:default-parms}
    \centering
    \begin{tabular}{ccc}
    Parameter & Default value & Physical meaning \\
    \hline\hline  
    $\alpha_\Lambda$ & 1.0 & turbulent length scale  \\
    $\alpha_\omega$   &  0.25  & non-locality \\
    $\alpha_\mathrm{s}$  &  $\frac{1}{2}\sqrt{\frac{2}{3}}$  & entropy flux \\
    C$_\mathrm{D}$ & $\frac{8}{3}\sqrt{\frac{2}{3}}$ & viscous dissipation \\
    $\gamma_\mathrm{R}$ & 2$\sqrt{3}$ & radiative dissipation \\
    $\beta$ & 1.0 & limitation of $\Lambda$ \\
    \hline
    \end{tabular}
\end{table}

The default parameters of the 1-equation model, as given in Table~\ref{tab:default-parms}, were estimated by comparing the convective flux and the convective velocity of the stationary, local limit of the 1-equation model with the results from MLT ($\alpha_\mathrm{s}$, $C_\mathrm{D}$), or by simple physical arguments ($\alpha_\omega$) \citep{Kuhfuss1986, Kuhfuss1987}. \citet{Wuchterl1998} suggested using the same value for $\gamma_\mathrm{R}$ as determined by \citet{Kuhfuss1986} for the 3-equation model, who determined it by comparison to MLT. 
For $\beta,$ we follow \citet{Straka2005}, who used a default value of $\beta=1$. 
The default value of $\alpha_\Lambda$ is motivated by the expectation that $\Lambda$ should be of the same order as the pressure scale height. However, this parameter can also be obtained from a solar calibration.
As $\alpha_\Lambda$ is always in combination with the other parameters, there are effectively five free parameters. 
The influence of the free parameters on the stellar structure is investigated in Sect.~\ref{sec:results-parameters}.

%%%%%%%%%%%%%%%%%%%%%%%%%%%%%%%%%%%%%%%%%%%%%%%%%%%%%%%%%%%%
\section{The Sun}\label{sec:sun}
%%%%%%%%%%%%%%%%%%%%%%%%%%%%%%%%%%%%%%%%%%%%%%%%%%%%%%%%%%%%

The vicinity of the Sun to the Earth enables us to observe and deduce several quantities that are directly linked to the convective envelope, and thus, can serve as reference points for solar models \citep[see e.g.][for a review]{Christensen-Dalsgaard2021}. 
In the following section, we describe the measurements available for comparison with the solar models, which are subsequently discussed.

\subsection{Observations}\label{sec:sun-observation}
%%%%%%%%%%%%%%%%%%%%%%%%%%%%%%%%%%%%%%%%%%%%%%%%%%%%%%%%%%%%%%%%%%%%%%%%%%%

Helioseismology is the study of the oscillations visible on the solar surface, which can be used to infer the interior structure and dynamics of the Sun \citep{Christensen-Dalsgaard2002}. 
It allows the radial sound speed and density profile of the Sun to be obtained with percentage accuracy or better \citep{Basu2009}. Furthermore, helioseismology allows the determination of the helium abundance in the envelope \citep[][$Y_\mathrm{cz}=0.2485\pm0.0034$]{Basu2004} and the location of the base of the convective zone \citep[][$R_\mathrm{cz} = 0.713 \pm 0.001$~R$_\odot$]{Basu1997b}, which is defined as the radius where the temperature gradient changes from an adiabatic to a radiative gradient \citep{Basu2016}. 
Depending on the prescription of CBM, this does not have to be coincident with the edge of the mixed region. 

The observed frequencies are also influenced by the detailed structure of the outermost, superadiabatic layers of the convective envelope. 
Inappropriate modelling of these layers and neglect of the non-adiabaticity of the oscillations \citep{Houdek2017} lead to a systematic difference between the observed and modelled frequencies. In helio- and asteroseismology, this is called the `surface effect'. We find that the 1-equation and MLT solar models have a very similar structure in the outer layers, leading to a very similar surface effect. Therefore, we do not discuss the surface effect in more detail in this work, but show the similarity in Appendix~\ref{ap:Se}. 
For a solution of the surface effect problem using three-dimensional box-in-a-star simulations, see \citet{Jorgensen2018b} and \citet{Jorgensen2019}. 

An additional and independent test of the quality of the 1-equation model concerns the central solar regions.
A 1~\Msun\ star is expected to develop a convective core at the end of the pre-MS, when the core temperature becomes sufficiently hot to bring the pp1 chain into $^3$He equilibrium and to burn $^{12}$C. When $^{12}$C is depleted and the pp chain dominates the energy production, this convective core should disappear again \citep{Kippenhahn2013}.
A convection theory has to reproduce this behaviour, and most importantly must result in a solar model that does not have a convective core at the age of the Sun. Models with an excessively efficient ad hoc overshooting implementation will result in a persistent convective core over the full duration of the solar MS lifetime.
For all models discussed in Sect.~\ref{sec:results}, we confirmed that the energy transport in the core at the age of the Sun is not convective. This validates the results of the previous successful test of the 1-equation Kuhfuss TCM for hydrogen-burning stellar cores \citep{Ahlborn2022}.

Another observable that is directly influenced by convection is the lithium abundance in the photosphere of stars. Li is already depleted at relatively low temperatures of about $2.5 \cdot 10^{6}$~K \citep{Pinsonneault1997}, for densities typically found in MS envelopes. The depletion of Li in convective envelopes depends on whether the well-mixed envelope reaches sufficiently deeply into the star, that is, into regions hot enough for Li depletion, and how long this stage lasts. 
Observations of open clusters and solar analogues suggest that Li depletion happens on the MS. 
However, simple convection models like MLT or the 1-equation model only predict Li depletion on the pre-MS. 
To reproduce the Li depletion on the MS, the convection theories need to be modified, for example by using MLT with different overshooting lengths for the pre-MS and MS evolution, and for core and envelope convection \citep{Schlattl1999}, or processes beyond convection need to be considered \citep[e.g. ][]{Andrassy2013, Montalban1994, Zahn1992, CaballeroNavarro2020}. 
Such additional processes are not included in the solar models discussed in this work, and therefore we do not discuss Li depletion in detail. We comment on this briefly in Appendix~\ref{ap:Li}.

\subsection{Standard solar models}\label{sec:sun-model}
%%%%%%%%%%%%%%%%%%%%%%%%%%%%%%%%%%%%%%%%%%%%%%%%%%%%%%%%%%%%%%%%%%%%%%%%%%%

%(see \todo{Serenelli+2016} for a review)
The GARching STellar Evolution Code \citep[GARSTEC, ][]{Weiss2008} calibrates a standard solar model (SSM) by matching a 1~\Msun\ model of the solar age to the Sun's present luminosity (\Lsun), radius (\Rsun), and surface metallicity fraction ($Z_\odot/X_\odot$) within 1 in $10^{5}$ parts \citep{Schlattl1997}.
The free parameters are the mixing length parameter ($\alpha_\mathrm{MLT}$), and the initial helium ($Y_\mathrm{init}$) and metal abundances ($Z_\mathrm{init}$). 
This calibration can also be applied to the 1-equation model, in which case the parameter $\alpha_\Lambda$ is varied, which is the analogue of the $\alpha_\mathrm{MLT}$ parameter in MLT.

The models include the atomic diffusion of hydrogen, helium, and metals in radiative regions. In convective regions, mixing of all elements is modelled by a diffusion equation with a diffusive velocity obtained from the convection theory. Effectively, this amounts to instantaneous mixing.
Solar models computed with MLT (Sect.~\ref{sec:results-mltvs1eq}) do not include overshooting. 

For a comparison between the models using MLT and the ones using the 1-equation model in Sect.~\ref{sec:results-mltvs1eq}, the parameters of the 1-equation model were fixed to the default parameters as determined by \citet{Kuhfuss1986} (see Table~\ref{tab:default-parms}), except for $\alpha_\Lambda$, which was varied to match the Sun. The abundances as determined by \citet{Magg2022} were used, with a metal fraction of $Z_\odot/X_\odot=0.0225$. 
In Sects.~\ref{sec:results-abundances} and \ref{sec:results-parameters}, solar calibrated models with different abundances and varied parameters of the 1-equation model were calculated to test the influence of these inputs. The abundances and parameters used are specified in the respective sections.

For all solar models calculated for this work, we adopted the opal equation of state from \citet{Rogers2002} and the OP opacities from \citet{Badnell2005}, substituted at low temperatures with the opacities by \citet{Ferguson2005}, using the appropriate compositions. 
For the stellar models discussed in Sect.~\ref{sec:results-mltvs1eq}, we used opacities calculated from the abundances determined by \citet{Magg2022} (Yago Herrera, private communication).
For the stellar models discussed in Sect.~\ref{sec:results-abundances} and \ref{sec:results-parameters}, we used the abundances as indicated in those respective sections.

%%%%%%%%%%%%%%%%%%%%%%%%%%%%%%%%%%%%%%%%%%%%%%%%%%%%%%%%%%%%
\section{Results}\label{sec:results}
%%%%%%%%%%%%%%%%%%%%%%%%%%%%%%%%%%%%%%%%%%%%%%%%%%%%%%%%%%%%

In the following section, we compare the models described in Sect.~\ref{sec:sun-model} to the solar seismic structure described in Sect.~\ref{sec:sun-observation}. First, we compare the solar calibrated model obtained using MLT (\mlt) with the one obtained using the Kuhfuss 1-equation model (\eq). For this comparison, the parameters of the 1-equation model were kept at the default values, except for the calibrated parameter $\alpha_\Lambda$. 
In Sect.~\ref{sec:results-abundances} we test the impact of different compositions on the structure of the solar model. We close this section by testing the influence of the parameters of the 1-equation model (Sect.~\ref{sec:results-parameters}).

\subsection{Comparing MLT and the 1-equation model}\label{sec:results-mltvs1eq}
%%%%%%%%%%%%%%%%%%%%%%%%%%%%%%%%%%%%%%%%%%%%%%%%%%%%%%%%%%%%%%%%%%%%%%%%%%%

%Convective boundary
The base of the convective envelope as measured by helioseismology is defined as the radius where the temperature gradient changes from close to adiabatic to radiative; this occurs at different radii depending on the convection theory used\footnote{see Appendix~\ref{ap:schematic-conv.zone} for a detailed description of the terminology for the different parts of the envelope.}. 
For the \mlt,\ this radius coincides with the Schwarzschild radius, where $\nabla_\mathrm{ad}=\nabla_\mathrm{rad}$, which equates to 0.7144~\Rsun\ (Table~\ref{tab:mltvs1eq}). This value is within $1.4\sigma$ of the seismologically inferred value of 0.713$\pm$0.001~\Rsun\ \citep{Basu1997b}, and is in agreement with the recent solar model by \citet{Magg2022}. %Magg2022 Rcz=0.7120Rsun
The 1-equation model predicts a nearly adiabatic temperature stratification over most of the CBM region. Thus, the change to $\nabla_\mathrm{rad}$ happens at the boundary of the CBM region, at a radius of 0.6845~\Rsun, which is in strong disagreement with the helioseismic measurement (29$\sigma$).

\begin{figure}
    \centering
    \includegraphics[width=\linewidth]{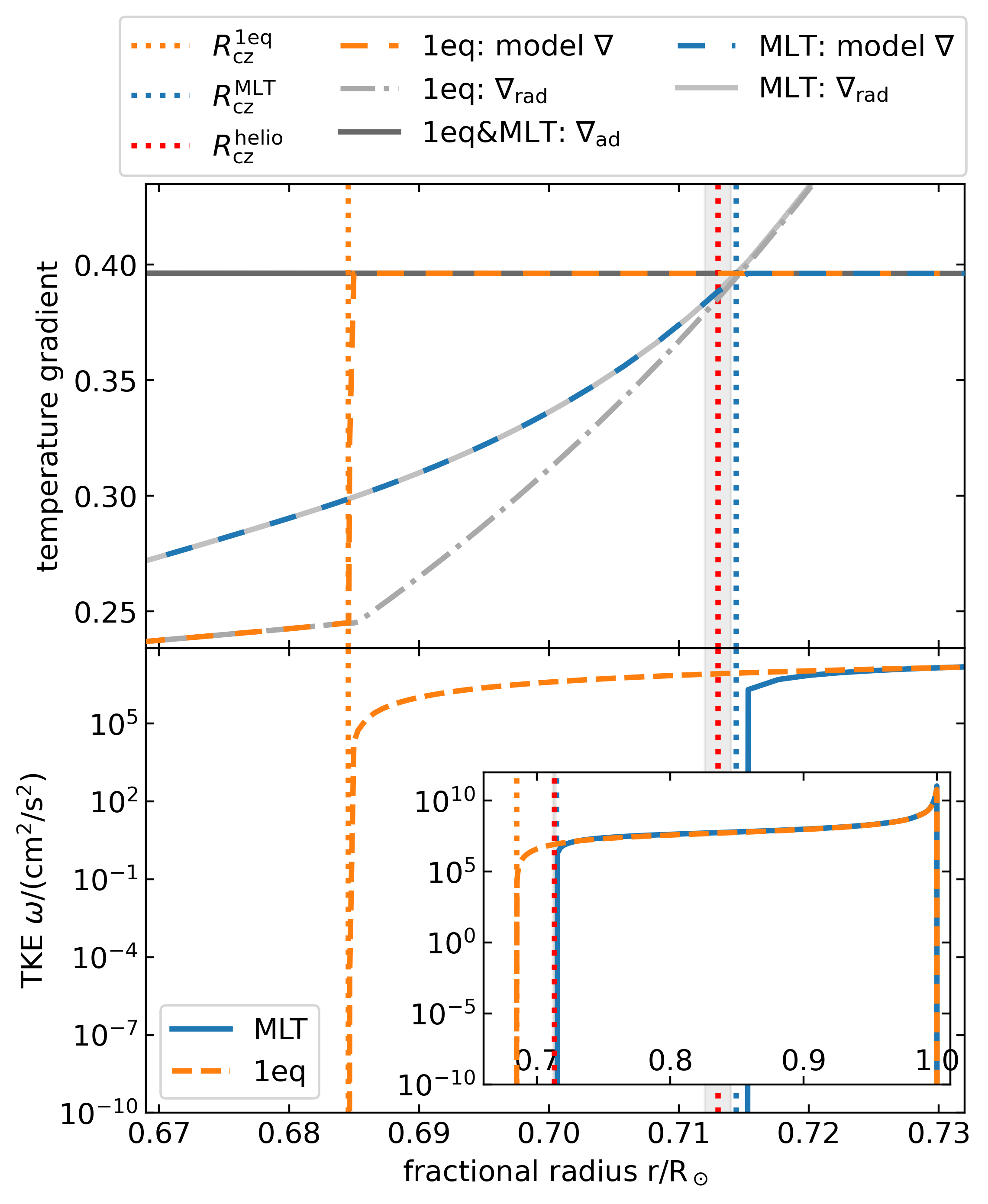}
    \caption{Temperature stratification (upper panel) and TKE $\omega$ (lower panel) in the region of the convective boundary, calculated with the 1-equation model (orange) and with MLT (blue), respectively.
    The adiabatic ($\nabla_\mathrm{ad}$, dark grey), radiative ($\nabla_\mathrm{rad}$: light grey, MLT: solid, 1-equation model: dash-dotted), and model temperature gradients are shown. $\nabla_\mathrm{ad}$ is the same for the 1-equation model and MLT.
    For the 1-equation model, $\nabla$ is slightly subadiabatic in the CBM region and sharply changes to $\nabla_\mathrm{rad}$ when $\omega$ (lower panel) drops to zero. The inset in the lower panel shows the complete convective envelope. The vertical lines mark the boundaries of the convective regions (orange: 1-equation model, blue: MLT) and the helioseismic measurement \citep[red, ][]{Basu1997b}, whereas the grey shaded regions denote the uncertainty of the measurement.}
    \label{fig:tempgradient}
\end{figure}

Figure \ref{fig:tempgradient} shows the temperature stratification close to the convective boundary for the \eq\ and the \mlt. 
The convective velocity in the \mlt, and therefore the TKE, drops to zero at the Schwarzschild boundary (see lower panel of Fig.~\ref{fig:tempgradient}). By construction, the Schwarzschild boundary is coincident with the boundary of the convective region when using MLT. At that radius, the temperature gradient also changes from the adiabatic to the radiative gradient.
For the 1-equation model, the region with a close-to-adiabatic gradient is extended to a smaller radius than expected from helioseismology, while the Schwarzschild boundary is still located within the observationally allowed range.
The temperature gradient $\nabla$ in the 1-equation model is dependent on the TKE $\omega$, $\nabla_\mathrm{ad}$, and $\nabla_\mathrm{rad}$ (see Eq. \ref{eq:tempgradient}); 
it is superadiabatic at radii larger than the Schwarzschild boundary, and in this region $\omega>0$ and $\nabla_\mathrm{ad}<\nabla_\mathrm{rad}$. In the CBM region, that is beyond the Schwarzschild boundary, the temperature gradient becomes subadiabatic ($\nabla-\nabla_\mathrm{ad}$ of the order of $-10^{-6}$). %$\omega$ is larger than zero at this radius, due to the non-locality of the model.
However, $\omega$ is large for the largest part of the CBM region, and thus convection is efficient and $\nabla$ is close to $\nabla_\mathrm{ad}$. Only at the boundary of the CBM region does $\omega$ sharply drop to zero.
When $\omega \rightarrow 0$, $\nabla \rightarrow \nabla_\mathrm{rad}$ (see Eq. \ref{eq:tempgradient}). This means that $\nabla$ stays close to adiabatic for most of the CBM region, and sharply changes to $\nabla_\mathrm{rad}$ at the boundary. This extended region where $\nabla$ is close to $\nabla_\mathrm{ad}$ in the CBM region is in disagreement with the measurement. 
The expression for the temperature gradient in the 1-equation model was determined with the use of the convective flux, which was estimated using the downgradient approximation (Eq.~\ref{eq:downgradientapprox_pi}).
This approximation is likely what causes the poor agreement between the temperature stratification of the model and the measurements of the Sun.

\begin{table}[htb]
\caption{Characteristics of the solar models and comparison to the measurements of $R_\mathrm{cz}$ and $Y_\mathrm{cz}$ \citep[][ respectively]{Basu1997b, Basu2004}}  
\label{tab:mltvs1eq}
    \centering
    \begin{tabular}{c|ccc}
    \hline\hline  
                                             &  MLT         &  1-eq.    &  Measurement \\
    \hline
    $R_\mathrm{cz}$ [\Rsun]                  &   0.7144     &  0.6845   & 0.713$\pm$0.001   \\
    $Y_\mathrm{cz}$                          &   0.2423     &  0.2456   & 0.2485$\pm$0.0034 \\
%    $X_\mathrm{Li, Surf}/X_\mathrm{Li, 0}$   &   0.204      &  $5\cdot10^{-5}$    &   0.007$\pm$0.001 \\
    $\alpha_\mathrm{MLT}$ or $\alpha_\Lambda$&   1.809      &  1.818    & $\cdots$ \\
    $Y_\mathrm{init}$                        &   0.2724     &  0.2702   & $\cdots$ \\
    $Z_\mathrm{init}$                        &   0.0184     &  0.0179   & $\cdots$ \\
    \hline
    \end{tabular}
\end{table}

\begin{figure}
    \centering
    \includegraphics[width=\linewidth]{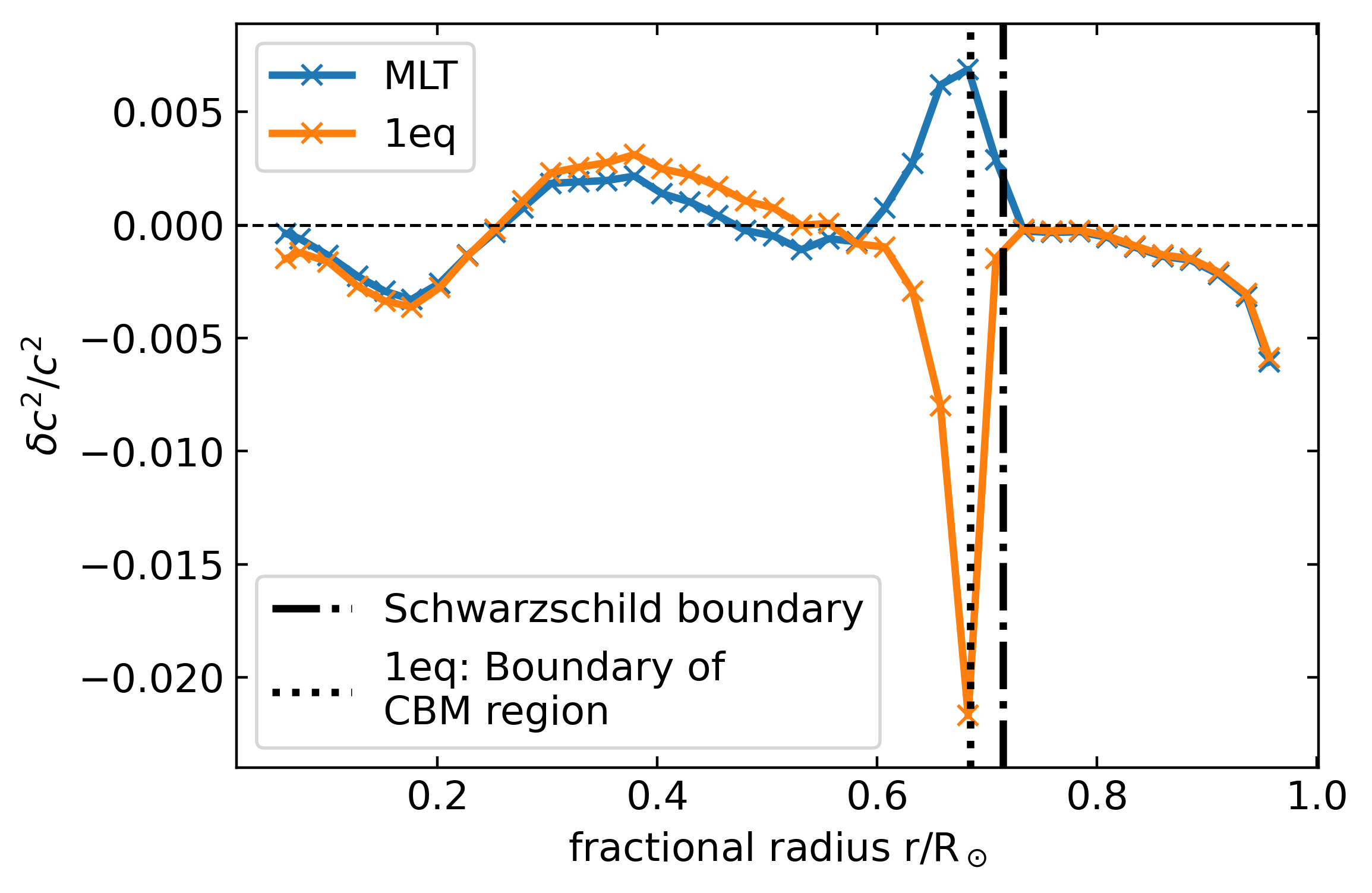}
    \caption{Sound speed profile as determined by \citet{Basu2009} compared to the profile predicted by \mlt\ (blue) and by \eq\ (orange). The y-axis shows the relative difference in the squared sound speed between the observation $c^2_\mathrm{helio}$ and the models $c^2_\mathrm{model}$: $\delta c^2/c^2 = (c^2_\mathrm{helio}-c^2_\mathrm{model})/c^2_\mathrm{helio}$}
    \label{fig:1eq-vs-mlt-csprofile}
\end{figure}

%sound speed profile
Figure \ref{fig:1eq-vs-mlt-csprofile} shows the relative difference in the squared sound speed between observations \citep{Basu2009} and solar models (\dcc).
The greatest deviation occurs for both models at a radius of $r\approx0.683$~\Rsun, in the vicinity of the base of the convective envelope. In this region, the \mlt\ underestimates the squared sound speed by $\approx 0.69$\%, whereas the \eq\ overestimates it by $\approx 2.17$\%. 
The larger sound speed at the base of the convective envelope obtained with the 1-equation model is connected to the location of the boundary of the CBM region \citep[see also][their Fig. 3]{Basu1997b}. 
In \eq, the change from a close-to-adiabatic to a radiative temperature gradient happens at a smaller radius than in the Sun.
Therefore, the predicted temperature in the CBM region is higher, resulting in a higher sound speed. 
The radii where the temperature gradient changes to a radiative gradient are marked with a vertical dash-dotted and dotted line for MLT and the 1-equation model, respectively.

%He abundance in the envelope
The He abundance in the envelope of both models is a good match to the seismically determined value (Table~\ref{tab:mltvs1eq}). For the \mlt,\ the abundance is 0.2423, which is within 1.8$\sigma$ of the measurement \citep[0.2485$\pm$0.0034,][]{Basu2004}, while it is 0.2456 for the \eq, within 0.9$\sigma$.
This small improvement is due to the larger convective reservoir, from which diffusive settling draws helium to the interior.

In summary, the agreement between observations and the solar model decreases when the Kuhfuss 1-equation model is used. This is especially obvious for quantities related to the temperature stratification, such as the boundary of the convective region and the sound speed profile. The He abundance is not affected. 
This is a hint that the downgradient approximation of the convective flux and the resulting temperature gradient do not appropriately represent the conditions in the Sun.

\subsection{Different abundances}\label{sec:results-abundances}
%%%%%%%%%%%%%%%%%%%%%%%%%%%%%%%%%%%%%%%%%%%%%%%%%%%%%%%%%%%%%%%%%%%%%%%%%%%5

The composition of the Sun is a long-standing problem. It is an important input parameter in solar models and influences, for example, the opacities, and thus observables such as the sound speed profile. 
In this section, we first give a brief overview of the commonly used compositions, and then we compare \eq\ with different compositions to investigate their impact on the model.

%GS98
\citet[][hereafter \citetalias{Grevesse1998}]{Grevesse1998} determined the fraction of metals in relation to hydrogen 
$Z_\odot/X_\odot$=0.023, using a 1D model of the solar atmosphere. 
%A09
With 3D hydrodynamic models becoming feasible, \citet{Asplund2005a} determined the composition of the Sun, which was updated by \citet[][hereafter \citetalias{Asplund2009}]{Asplund2009}. Their work resulted in a shift of the metallicity to lower values compared to \citetalias{Grevesse1998}: $Z_\odot/X_\odot$=0.0181.
%A09 discussion
The lower abundances derived by \citetalias{Asplund2009}, especially those of C, N, and O, result in lower opacities, and decrease the agreement between observations and a SSM calculated with the abundances from \citetalias{Asplund2009}. 
For a discussion of the possible causes and solutions for the decreased agreement, we refer the reader to \citetalias{Asplund2009}, \citet{Serenelli2016}, and \citet{Buldgen2019b}.
%M22
\citet[][hereafter \citetalias{Magg2022}]{Magg2022} redetermined the solar abundances, also using 3D simulations, but employing different atomic data. They obtained an overall metallicity fraction of $Z_\odot/X_\odot$=0.0225, the value used in Sect.~\ref{sec:results-mltvs1eq}. This is close to that determined by \citetalias{Grevesse1998}, but the abundances of individual elements differ. 
This improves the agreement with helioseismic measurements compared to the lower metallicity obtained by \citetalias{Asplund2009} \citep{Magg2022}. 

\begin{figure}
    \centering
    \includegraphics[width=\linewidth]{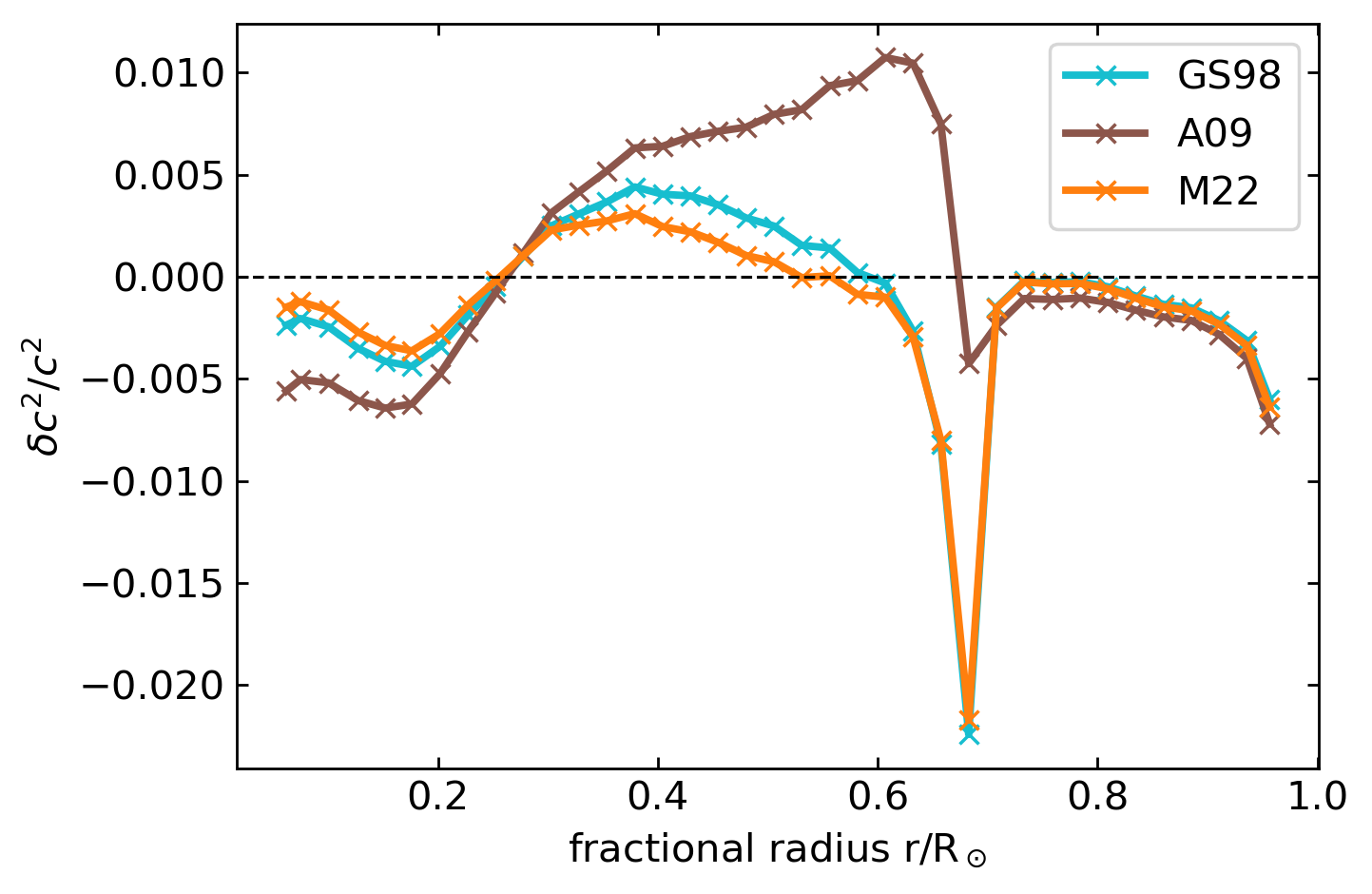}
    \caption{As in Fig.~\ref{fig:1eq-vs-mlt-csprofile}, but comparing the sound speed profiles of solar models calculated with the 1-equation model and different abundances: \citet[][\citetalias{Grevesse1998}]{Grevesse1998}, \citet[][\citetalias{Asplund2009}]{Asplund2009}, and \citet[][\citetalias{Magg2022}]{Magg2022}. The quantity on the y-axis is defined as $\delta c^2/c^2 = (c^2_\mathrm{helio}-c^2_\mathrm{model})/c^2_\mathrm{helio}$.
    }
    \label{fig:csprofile-abundances}
\end{figure}

%summary
However, the question of which abundances are the best to use is not straightforward.
For example, \citet{Buldgen2024a} studied the first adiabatic exponent using helioseismic inversions, and found a better agreement for lower metallicities. In a detailed study of the degeneracies and the influence of the input physics of a SSM, \citet{Buldgen2024b} found that the higher metallicities determined by \citetalias{Magg2022} do not solve the issues present in solar modelling.
Therefore, it is not yet clear which set of abundances is the appropriate one, and we calculated SSMs with the 1-equation model with all three sets of abundances.
%
%our results
The other input parameters are the same as described in Sect.~\ref{sec:sun-model}.
The results of these calculations are given in Table~\ref{tab:abundnaces} and Fig.~\ref{fig:csprofile-abundances}. 
We emphasise that in each case, we used opacities consistent with the solar composition.

The results obtained for the abundances from \citetalias{Magg2022} and \citetalias{Grevesse1998} are quite similar due to the similar total metallicity. 
The base of the convective zone is within 29$\sigma$ of the helioseimsic measurement for both \citetalias{Grevesse1998} and \citetalias{Magg2022}; it is improved, although still very different from the measurement (19$\sigma$), when using the abundances from \citetalias{Asplund2009}.

Figure~\ref{fig:csprofile-abundances} shows \dcc\ for the models with different abundances. 
The differences range from -2.24\% to 0.44\% and from -2.17\% to 0.31\% for \citetalias{Grevesse1998} and \citetalias{Magg2022}, respectively, and from -0.70\% to 1.07\%  for \citetalias{Asplund2009}. The morphology of \dcc\ is different between the models using the \citetalias{Asplund2009}-abundances and the ones using the \citetalias{Grevesse1998} and \citetalias{Magg2022} abundances. The largest absolute deviation in the latter two sets of abundances is just below the convective envelope and negative in sign, because the models over-predict the sound speed. For the \citetalias{Asplund2009} model, the maximum deviation is at smaller radii and the model under-predicts the sound speed.

\citetalias{Grevesse1998} and \citetalias{Magg2022} agree with the measured helium abundance in the envelope within 2$\sigma$ and 0.9$\sigma$, respectively. For \citetalias{Asplund2009}, the agreement is decreased to 3.3$\sigma$.

As was found previously for MLT-based SSMs \citep{Magg2022}, the results for \citetalias{Grevesse1998} and \citetalias{Magg2022} are nearly identical. For our TCM-based models, 
the agreement between the helioseismic measurements and \citetalias{Asplund2009} is better for the depth of the convective envelope compared to the models with the other abundances. However, it is worse when considering the He content in the solar envelope, and the sound speed profile.

\begin{table}[htb]
\caption{Characteristics of the solar models calculated with different abundances and comparison to the measurements of $R_\mathrm{cz}$ and $Y_\mathrm{cz}$ \citep[][ respectively]{Basu1997b, Basu2004}}  
\label{tab:abundnaces}
    \centering
    \begin{tabular}{c|cccc}
    \hline\hline  
                              &  \citetalias{Grevesse1998}    &  \citetalias{Asplund2009}   & \citetalias{Magg2022}      &  Measurement    \\
    \hline
    $R_\mathrm{cz}$ [\Rsun]                &   0.6842             &  0.6938           &   0.6845           &   0.713$\pm$0.001   \\
    $Y_\mathrm{cz}$                        &   0.2416             &  0.2371           &   0.2456           &   0.2485$\pm$0.0034 \\
    $\alpha_\Lambda$                       &   1.817              &  1.808            &   1.818            &   $\cdots$ \\
    $Y_\mathrm{init}$                      &   0.2658             &  0.2623           &   0.2702           &   $\cdots$ \\
    $Z_\mathrm{init}$                      &   0.0184             &  0.0146           &   0.0179           &   $\cdots$ \\
    \hline
    \end{tabular}
\end{table}

\subsection{Varying the parameters of the 1-equation model}\label{sec:results-parameters}
%%%%%%%%%%%%%%%%%%%%%%%%%%%%%%%%%%%%%%%%%%%%%%%%%%%%%%%%%%%%%%%%%%%%%%%%%%%

\begin{figure*}[htb!]
    \centering
    \includegraphics[width=\linewidth]{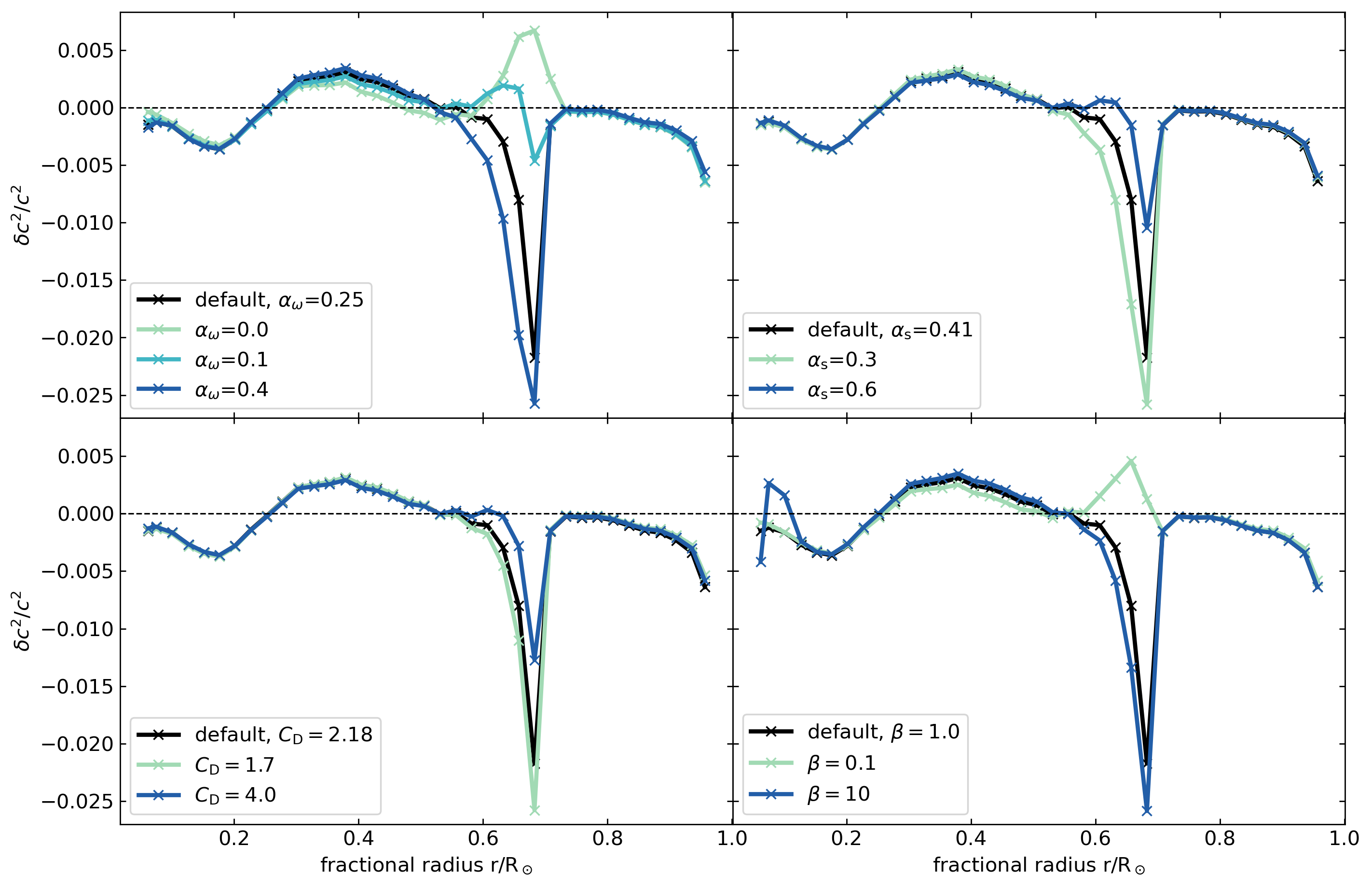}
    \caption{
    As in Fig.~\ref{fig:1eq-vs-mlt-csprofile}, but comparing the sound speed profiles of solar models calculated with the 1-equation model and different values for the free parameters. One parameter is varied in each panel while keeping the other parameters at the default values. The parameter changed is specified in the legend. The quantity on the y-axis is defined as $\delta c^2/c^2 = (c^2_\mathrm{helio}-c^2_\mathrm{model})/c^2_\mathrm{helio}$.}
    \label{fig:1eq-parms-csprofile}
\end{figure*}

The Kuhfuss 1-equation model effectively has five free parameters (see Sect.~\ref{sec:kuhfuss}). 
\citet[see their Appendix]{Ahlborn2022} investigated the influence of some parameters for the cores of MS models, and found a rather limited effect. 
Here we test their influence on the solar convective envelope. We performed a solar calibration, varying one of the parameters at a time, while leaving the others at their default values. For these calculations, the abundances determined by \citet{Magg2022} were used. The results of the solar calibration with different sets of parameters are shown in Fig.~\ref{fig:1eq-parms-csprofile} and Table~\ref{tab:parmgrid}. For each run, the varied parameter and its new value are given in the header of Table~\ref{tab:parmgrid}, and all other parameters except for the specified one were kept at the default value (see Table~\ref{tab:default-parms}).
As $\alpha_\Lambda$, $Y_\mathrm{init}$, and $Z_\mathrm{init}$ are varied to match \Rsun, \Lsun, and $Z_\odot/X_\odot$ in the solar calibration, the values of $\alpha_\Lambda$, $Y_\mathrm{init}$, and $Z_\mathrm{init}$ differ between calculations with different sets of parameters, partly compensating for the variation of the parameters of the TCM.

We find that $R_\mathrm{cz}$ is at a smaller radius, and that the higher the value of $\alpha_\omega$, the lower the values of $\alpha_\mathrm{s}$ and C$_\mathrm{D}$. This also affects the other observables: When the convective envelope reaches deeper into the solar model, the value of \dcc\ in the vicinity of the boundary of the CBM region is more negative (Fig.~\ref{fig:1eq-parms-csprofile}), and the He abundance in the envelope is higher (Table~\ref{tab:parmgrid}).
These effects can be understood by the physical interpretation of the individual free parameters. 

The non-locality of the model is controlled by $\alpha_\omega$. Hence, an increase in $\alpha_\omega$ results in an extension of the CBM region. 
The parameter $\alpha_\mathrm{s}$ combined with $\Lambda$ and $\omega^{1/2}$ can be interpreted as a diffusion coefficient for the convective flux, which is powered by the entropy gradient (see Eq.~\ref{eq:downgradientapprox_pi}). Thus, decreasing $\alpha_\mathrm{s}$ decreases the convective flux for the same entropy gradient. In other words, to obtain the same convective flux, which is needed to reproduce \Lsun\ and \Rsun, the product $\Lambda \omega^{1/2}$ needs to increase when $\alpha_\mathrm{s}$ is decreased. It was confirmed that the quantities $\Lambda$ ---which depends on $\alpha_\Lambda$--- and $\omega$ increase for decreased $\alpha_\mathrm{s}$. %note: the superadiabatic gradient is not changed when alpha_s is changed.
As the non-local term increases for a generally larger $\omega$ and $\Lambda$ (see Eq.~\ref{eq:non-local_effects}), the CBM region becomes more extended when decreasing $\alpha_\mathrm{s}$.

Finally, the viscous dissipation is controlled by C$_\mathrm{D}$. Thus, a smaller value for C$_\mathrm{D}$ decreases the dissipation in the convective region, increasing the efficiency of the convection, which means increasing $\omega$. Again, this increase in $\omega$ also effects the CBM region, shifting the boundary of the CBM region to smaller radii.

When varying $\gamma_\mathrm{R}$, the parameter controlling the radiative dissipation, only minor changes are visible. Thus, the exact value of the parameter of the radiative dissipation does not significantly  influence the result.

The parameter $\beta$ is not a parameter that directly controls the physical processes playing a role in convection, but instead it was introduced to limit the turbulent length scale $\Lambda$ in regions where the pressure scale height $H_\mathrm{p}$ diverges \citep{Wuchterl1995}. The larger the correction applied to $\Lambda$, the smaller the value of $\beta$.
The pressure scale height diverges in the centre, and therefore the limitation of $\Lambda$ due to $\beta$ mostly effects the core. 
For $\beta=10,$ it can be observed that the limitation is not strong enough in this case. The high values of $H_\mathrm{p}$ in the core region influence $\Lambda$ enough to alter the solar sound speed profile, which shows the need for a limitation of $\Lambda$ when $H_\mathrm{p}$ diverges (Fig.~\ref{fig:1eq-parms-csprofile}). 
However, $\beta$ also limits $\Lambda$ in the envelope. The impact on the convective envelope can be seen from Eq.~(\ref{eq:1eq}), where $\Lambda$ appears in the numerator of the source terms and in the denominator of the sink terms. Thus, if $\Lambda$ is limited less strongly by a larger $\beta$, $\omega$ has to be larger to compensate, and the convective envelope reaches deeper into the Sun due to the influence of $\omega$ on the non-local term.
As for the other cases, a deeper convective envelope (larger $\beta$) produces a sound speed profile of the model that tends to overestimate the sound speed near the boundary of the CBM region, resulting in a negative amplitude in Fig.~\ref{fig:1eq-parms-csprofile}. 

Therefore, varying the free parameters in the 1-equation model would allow the 1-equation model to be tuned such that the agreement between \eq\ and the helioseismic measurements is improved. However, this is not the aim of the present paper, and would not address the shortcomings of the 1-equation model (see Sect.~\ref{sec:discussion}).

\begin{table*}[htb]
\caption{Characteristics of the solar models calculated with different Kuhfuss parameters and comparison to the measurements of $R_\mathrm{cz}$ and $Y_\mathrm{cz}$ \citep[][ respectively]{Basu1997b, Basu2004}}
\label{tab:parmgrid}
    \centering
    \begin{tabular}{c|ccccccccc}
    \hline\hline  
                       &  default   &  $\alpha_\omega = 0.0$ &  $\alpha_\omega = 0.1$&  $\alpha_\omega = 0.4$&  $\alpha_\mathrm{s} = 0.3$&  $\alpha_\mathrm{s} = 0.6$&  $C_\mathrm{D} = 1.7$&  $C_\mathrm{D} = 4.0$  &  Measurement\\
    \hline
    $R_\mathrm{cz}$ [\Rsun]&  0.6845  & 0.7142 & 0.6957 & 0.6762 & 0.6779 & 0.6911 & 0.6822 & 0.6896   & 0.713$\pm$0.001 \\
    $Y_\mathrm{cz}$        &   0.2456 & 0.2424 & 0.2444 & 0.2464 & 0.2461 & 0.2449 & 0.2456 & 0.2451   & 0.2485$\pm$0.0034 \\
    $\alpha_\Lambda$       & 1.818    & 1.728  & 1.761  & 1.873  & 2.346  & 1.342  & 1.730  & 2.068    & $\cdots$  \\
    $Y_\mathrm{init}$      & 0.2702   & 0.2723 & 0.2708 & 0.2699 & 0.2698 & 0.2705 & 0.2698 & 0.2705   & $\cdots$  \\
    $Z_\mathrm{init}$      & 0.0179   & 0.0184 & 0.0180 & 0.0178 & 0.0178 & 0.0180 & 0.0178 & 0.0179   & $\cdots$  \\
    \hline 
    \end{tabular}
    \centering
    \begin{tabular}{c|cccccc}
    \hline\hline  
                       &  default   &  $\gamma_\mathrm{R} = 1.0$ &  $\gamma_\mathrm{R} = 5.5$&  $\beta = 0.1$&  $\beta = 10$  &  Measurement\\
    \hline
    $R_\mathrm{cz}$ [\Rsun]   &  0.6845  & 0.6851 & 0.6840 & 0.7019 & 0.6803  & 0.713$\pm$0.001 \\
    $Y_\mathrm{cz}$           &   0.2456 & 0.2455  & 0.2455 & 0.2438 & 0.2463 & 0.2485$\pm$0.0034 \\
    $\alpha_\Lambda$          & 1.818    & 1.780  & 1.860  & 1.829  & 1.817   & $\cdots$ \\
    $Y_\mathrm{init}$         & 0.2702   & 0.2702 & 0.2701 & 0.2712 & 0.2703  & $\cdots$ \\
    $Z_\mathrm{init}$         & 0.0179   & 0.0179 & 0.0179 & 0.0181 & 0.0178  & $\cdots$ \\
    \hline
    \end{tabular}
\end{table*}

%%%%%%%%%%%%%%%%%%%%%%%%%%%%%%%%%%%%%%%%%%%%%%%%%%%%%%%%%%%%
\section{Discussion}\label{sec:discussion}
%%%%%%%%%%%%%%%%%%%%%%%%%%%%%%%%%%%%%%%%%%%%%%%%%%%%%%%%%%%%

%papers from Xiong+; JCD; Petri; Schlattl,Weiss,Ludwig; Joergensen,Weiss   \\
There are several studies in the literature that investigate the thermal structure of the CBM in general and also specifically for the Sun. \citet{Xiong2001a} and \citet{Christensen-Dalsgaard2011} both used 1D stellar models to study the solar convective envelope, but followed different approaches. \citet{Christensen-Dalsgaard2011} investigated which parameterisation of the temperature stratification in the CBM region best  fits the helioseismic measurements. 
The theory used by \citet{Xiong2001a} is a non-local convection model based on the auto- and cross-correlations of the temperature fluctuations and the turbulent velocity (\citealp{Xiong1989}; see \citealp{Xiong2021} for a review of this theory). This theory is similar to the full Kuhfuss 3-equation model, where the correlations of the turbulent velocity and the entropy fluctuations are used. Both \citet{Xiong2001a} and \citet{Christensen-Dalsgaard2011} found a smooth transition of the temperature gradient from close-to-adiabatic to radiative in the CBM region, and an extended subadiabatic layer already within the formal Schwarzschild boundary. Using the non-local and anisotropic model derived by \citet{Xiong1989} and \citet{Deng2006}, \citet{Zhang2012b} constructed a model for the solar envelope, and showed that the discrepancies in the sound speed profile decrease if a non-local convection model with such a smooth transition of the temperature gradient is used.

\citet{Kapyla2017} employed a 3D hydrodynamic simulation to study the CBM region, finding that a convective region is generally composed of three layers: a buoyancy region with a super-adiabatic gradient and positive convective flux, followed by a Deardorff layer, which is a sub-adiabatic layer with positive convective flux, and an overshooting region, where the convective flux becomes negative and the temperature gradient is subadiabatic. The overshooting region is then followed by a region where radiative energy transport is dominating. These layers have the same characteristics as the ones found by \citet{Christensen-Dalsgaard2011} and \citet{Xiong2001a} using one-dimensional approaches.
\citet{Kapyla2017} confirmed that the convective flux has a contribution that is not proportional to the entropy gradient, which is found to be crucial for developing the Deardorff layer \citep[see also ][]{Deardorff1961, Deardorff1966}. 
The 1-equation model, as well as MLT, use the downgradient approximation for the convective flux, which means that the convective flux is modelled by assuming a proportionality with the entropy gradient (Eq.~\ref{eq:downgradientapprox_pi}). This assumption prevents a Deardorff layer, and causes a physically inaccurate modelling of the CBM region.

Using 2D hydrodynamic simulations, \citet{Baraffe2021} found that CBM can modify the thermal background in the CBM region considerably. This leads to a more gradual change of the temperature gradient in the CBM region. 
\citet{Baraffe2022} included the description of the modification of the temperature gradient found by \citet{Baraffe2021} in a 1D stellar evolution code, finding that this effect has the potential to improve the sound speed discrepancy, which is found just below the convective envelope in solar models.

All the aforementioned findings show that the temperature stratification is not appropriately modelled in the 1-equation model, which is connected to the approximation of the convective flux based on the entropy gradient.
As briefly mentioned in Sect.~\ref{sec:kuhfuss}, this approximation is the main difference between the Kuhfuss 1-equation and 3-equation models. In the 3-equation model, the equation for the convective flux variable $\Pi$ is kept. 
With the temperature gradient given as 
\begin{equation}
    \nabla = \nabla_\mathrm{rad} - \frac{H_\mathrm{p}}{k_\mathrm{rad} T} \rho T \Pi \, ,
\end{equation}
it becomes clear that keeping the equation for $\Pi$ without employing the downgradient approximation will have a direct impact on the temperature stratification of the model. 
In earlier studies, the 3-equation model was applied to convective cores, and it was confirmed that the 3-equation model is able to produce not only the correct convective core size for MS stars, but also a Deardorff layer and a more gradual change of the temperature gradient at the boundary of the CBM region \citep{Ahlborn2022}.
The next step in this project will therefore be to apply the 3-equation model in stellar envelopes and to test this again on a solar model. Due to the findings in main sequence stars, we expect the 3-equation model to provide a better structure of the solar CBM region.

In standard solar models, several physical effects are not included. Their purpose is to serve as a well-defined reference against which new physical ingredients can be tested \citep[see][for a review]{Serenelli2016}. 
However, an appropriate modelling of the convective energy transport is not the only open question that needs to be answered in order to find a realistic solar model. 
The depletion of Li is a prime example of possibly missing physics, and is needed to bring the depletion of Li as a function of time into agreement with observations. In Appendix~\ref{ap:Li}, we briefly show the evolution of the solar surface Li content as a result of our \eq\ model.
Other physical processes known to influence the Sun but nevertheless not included in standard solar models are magnetic fields \citep{Baldner2009} and rotation \citep{Thompson2003}. It is likely that the appropriate treatment of these processes necessitates sophisticated (magneto-)hydrodynamic simulations.

%%%%%%%%%%%%%%%%%%%%%%%%%%%%%%%%%%%%%%%%%%%%%%%%%%%%%%%%%%%%
\section{Conclusion and summary}\label{sec:conclusion+summary}
%%%%%%%%%%%%%%%%%%%%%%%%%%%%%%%%%%%%%%%%%%%%%%%%%%%%%%%%%%%%

We used a standard solar model to test the 1-equation convection model first developed by \citet{Kuhfuss1986}.
The advantage of this TCM is that the non-locality of convection is included in the theory and no ad hoc inclusion of CBM is needed.
Comparison with several observed quantities of the Sun shows that the convective envelope as predicted by a solar-calibrated model with the 1-equation Kuhfuss model reaches too far into the star if the default choice of parameters suggested by \citet{Kuhfuss1986} is used. While substantial CBM from a convective envelope would be a way to solve the red giant bump problem (see Sect.~\ref{sec:intro}), it would degrade the SSM quality, at least when the temperature gradient is nearly adiabatic in the CBM region.

The nearly adiabatic temperature stratification in the CBM region is connected to the simplicity of the Kuhfuss 1-equation model. The close coupling of the convective flux and the super-adiabatic temperature gradient ---which is introduced by the downgradient approximation--- results in an unrealistic temperature stratification in the CBM region. Several 2D and 3D simulations, as well as other 1D models \citep{Christensen-Dalsgaard2011, Xiong2001a, Kapyla2017, Baraffe2021}, found that the temperature gradient in the CBM region is subadiabatic, smoothly changing to a radiative gradient; whereas in the 1-equation model, it is close to adiabatic and changes to a radiative gradient abruptly at the boundary of the CBM region. 
The 3-equation model, the full version of the Kuhfuss-TCM \citep{Kuhfuss1987, Kupka2022}, lifts the tight coupling between the convective flux and the superadiabatic temperature gradient. \citet{Ahlborn2022} applied the 3-equation model to convective cores and found a temperature stratification in better agreement with other TCMs and hydrodynamical simulations. This is a first suggestion that the modelling of the temperature stratification is more realistic, and that a solar model can be improved by using the 3-equation model. This will be investigated further in the future.

Another route to a better solar model could be to calibrate the free parameters of the Kuhfuss 1-equation model. Their default values were mostly determined by calibrating the local and stationary limit of the 1-equation model to the MLT, but this is not compulsory. Instead, the helioseismic measurements could be used to calibrate the free parameters. However, considering the shortcomings of the 1-equation model, this approach does not seem to be very favourable, and a convection model with a more realistic temperature stratification would be a more promising way forward. 

The calibration of the Kuhfuss 1-equation model to the MLT in the local and stationary limit results in a similar structure in the superadiabatic parts of the solar envelope (Fig.~\ref{fig:envstruc}). This, and the failure of the 1-equation TCM concerning the depth of the solar convective envelope, indicate that the Kuhfuss 1-equation model can be considered a consistent and physical non-local extension of MLT. It may be used instead of the normal approach of extending MLT with ad hoc overshooting. Indeed, first tests demonstrate that the mentioned red giant bump problem can be solved with the 1-equation Kuhfuss model.  Why the theory seems to work for deep convective envelopes in red giants but overpredicts the depth of the comparably shallow solar convective envelope remains to be understood.

On the other hand, in spite of the non-locality of the 1-equation model, the convective core that exists during the final pre-MS phase vanishes before reaching the MS, such that the solar model has no convective core. This supports the conclusion of \citet{Ahlborn2022} that the 1-equation model is a reasonable theory for core convection.

\begin{acknowledgements}
We thank Yago Herrera for providing the opacity tables with the updated solar compositions. We thank Andreas J{\o}rgensen for providing the data of the patched solar model, and for helpful discussions. The research leading to the presented results has received funding from the European Research Council under the European Community’s Horizon 2020 Framework/ERC grant agreement no 101000296 (DipolarSounds).
\end{acknowledgements}

\bibliographystyle{aa} % style aa.bst
\bibliography{bib-sun, bib-convgeneral, bib-kippenhahn}

\begin{appendix}

\section{Envelope structure}\label{ap:Se}

\begin{figure}
    \centering
    \includegraphics[width=\linewidth]{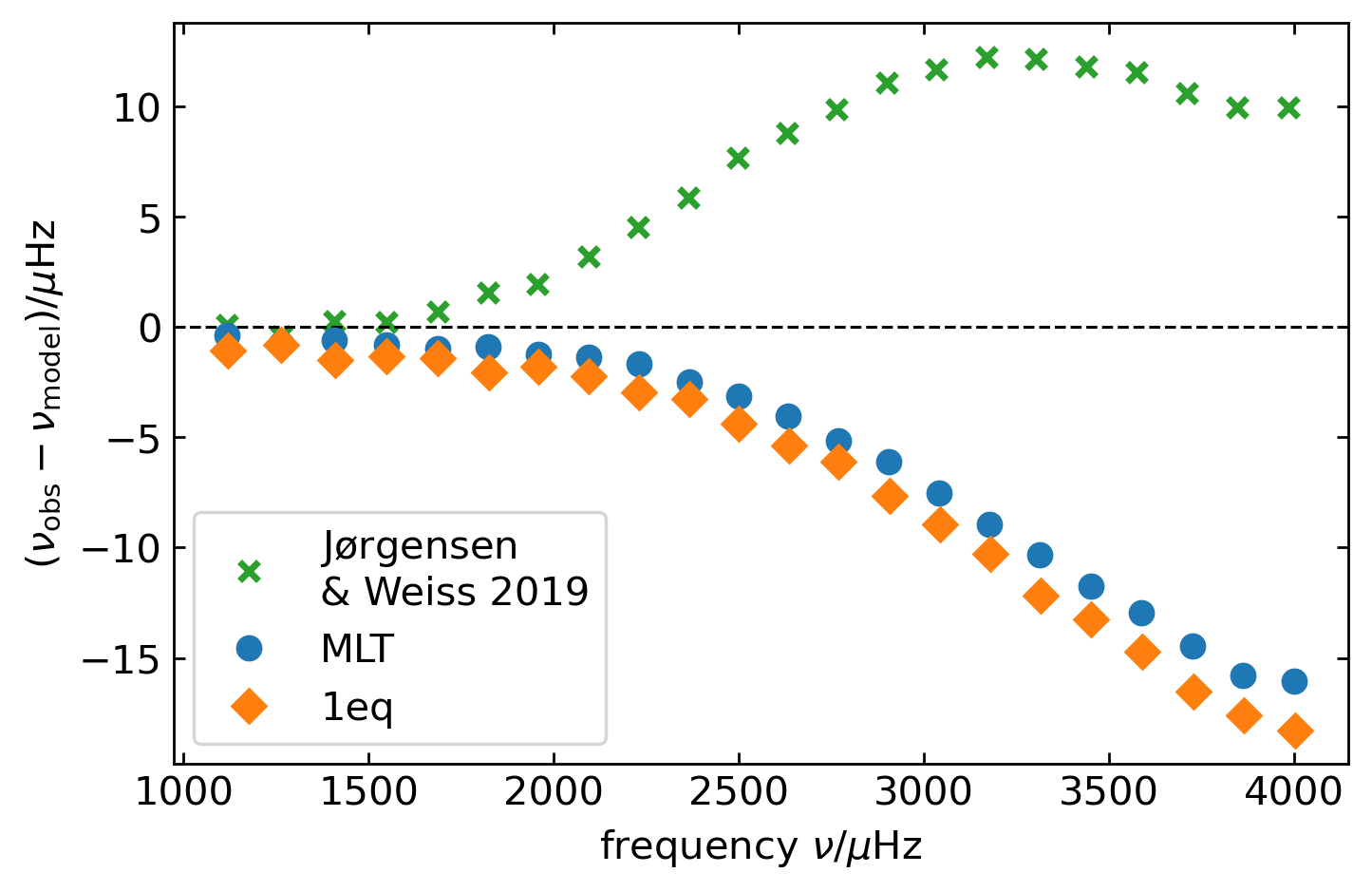}
    \caption{The difference between the observed frequencies of the $l=0$ modes \citep{Broomhall2009, Davies2014}, and the frequencies calculated based on the \eq\ (orange diamonds), \mlt\ (blue circles), and the solar model from \citet{Jorgensen2019}, with inclusion of turbulent pressure and patched 3D atmosphere (green crosses) is shown. For the solar model from \citet{Jorgensen2019}, the structural surface effect is corrected, and the difference between observed and calculated frequencies can be explained by remaining modal effects.}
    \label{fig:surfeff}
\end{figure}

\begin{figure}
    \centering
    \includegraphics[width=\linewidth]{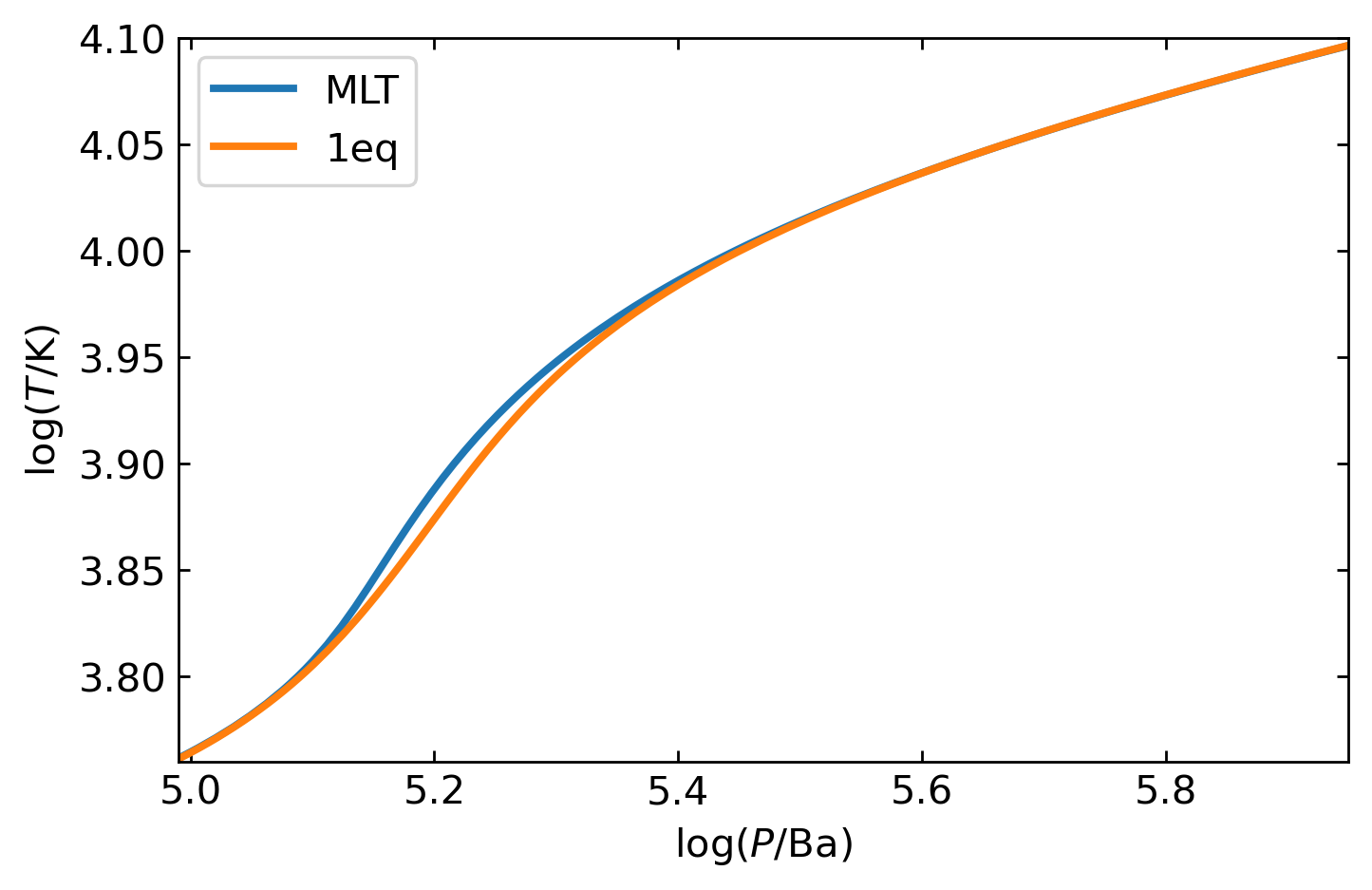}
    \caption{The structure of the envelope of the \eq\ (orange) and of the \mlt\ (blue) is very similar, which leads to a very similar surface effect.}
    \label{fig:envstruc}
\end{figure}

As mentioned in Sect.~\ref{sec:sun}, SSM employing MLT predict a defective thermal structure of the convective envelope that results in the surface effect, a systematic deviation of model p-mode frequencies from those observed in the Sun. In Fig.~\ref{fig:surfeff} we show the comparison of $l=0$ p-mode frequencies resulting from the \mlt, the \eq, and a SSM with a patched envelope and atmosphere from a three-dimensional simulation \citep{Jorgensen2019}. Although in the latter there is still a discrepancy visible, the authors argue that this is due to the adiabaticity assumed in the frequency calculation (the ``modal effect''), while the ``structural effect'' is solved. In any case, it is clear that both MLT and the 1-equation Kuhfuss model share the same problem.
This is the result of the very similar thermal structure in both models, as is shown in Fig.~\ref{fig:envstruc}.

\section{Li depletion} \label{ap:Li}

\begin{figure}
    \centering
    \includegraphics[width=\linewidth]{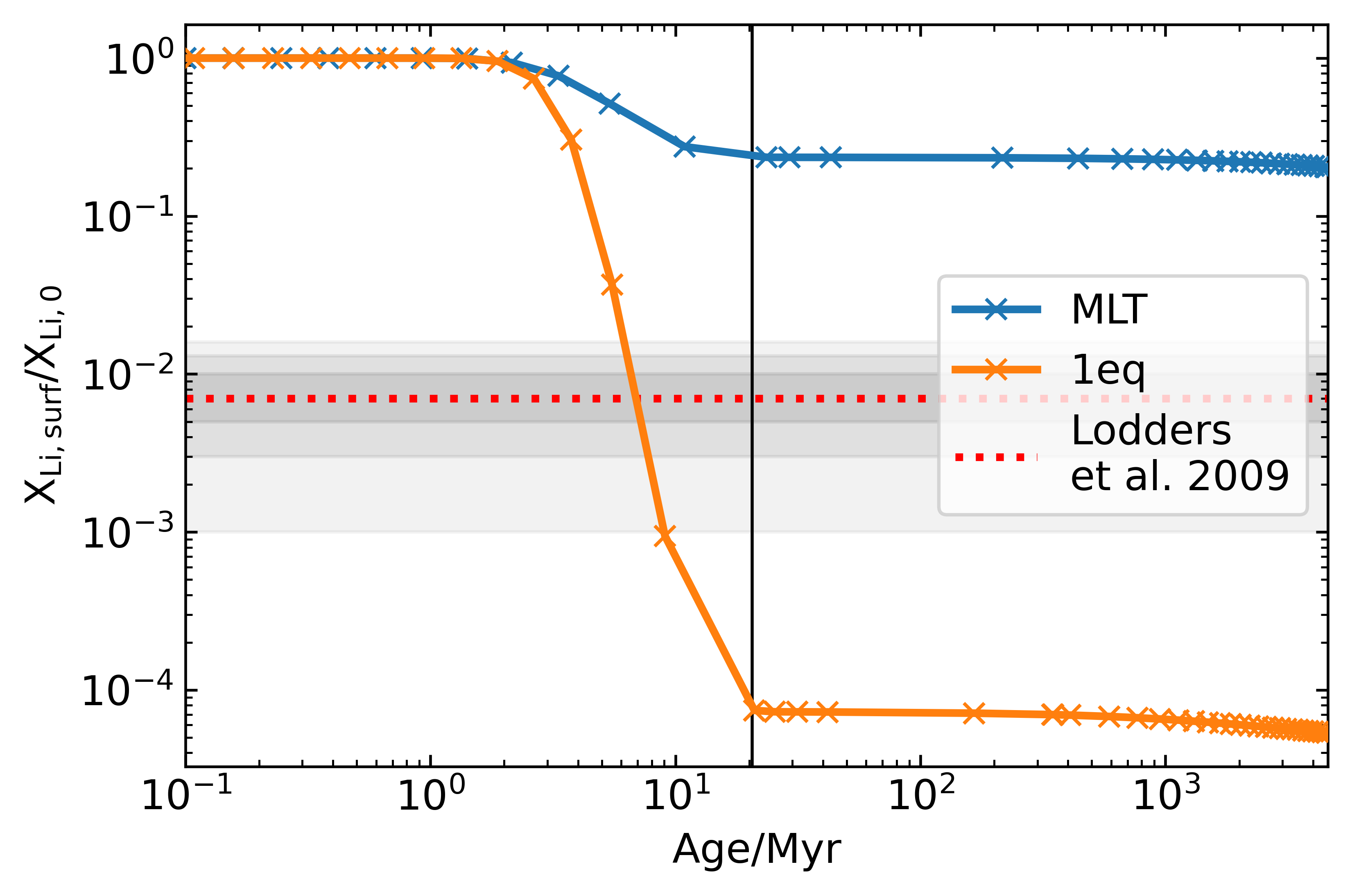}
    \caption{Li is depleted over the lifetime of the Sun. For comparison, the value derived from meteoric and photometric measurements is included as red dashed line, with the 1, 2 and 3$\sigma$ range denoted as grey shaded regions \citep{Lodders2009}. The vertical black line denotes the beginning of the MS.}
    \label{fig:li}
\end{figure}

\begin{table}[htb]
\caption{Comparison of the Li depletion in the solar models and the measurement \citep{Lodders2009}}  
\label{tab:Limltvs1eq}
    \centering
    \renewcommand{\arraystretch}{1.5}
    \begin{tabular}{c|ccc}
    \hline\hline  
                                             &  MLT         &  1-eq.    &  Measurement \\
    \hline
    $X_\mathrm{Li, Surf}/X_\mathrm{Li, 0}$   &   0.204      &  $5\cdot10^{-5}$    &   0.007$^{+0.003}_{-0.002}$ \\
    \hline
    \end{tabular}
\end{table}

%Li depletion
Figure \ref{fig:li} shows the depletion of lithium as a function of age as predicted by the \mlt\ and the \eq\ (Table~\ref{tab:Limltvs1eq}). 
The fractional Li depletion as determined by \citet{Lodders2009} is used as a reference for the expected depletion, and is marked in the figure with a red dotted line, with grey shaded areas to indicate the 1$\sigma$, 2$\sigma$ and 3$\sigma$ ranges. 
While it is clear that the \mlt\ has a fractional Li abundance which is much too high, the fractional Li abundance as predicted by the \eq\ is in better agreement (3.5$\sigma$), but lower than the observation. 
The extension of the well-mixed CBM region when using the 1-equation model brings Li to smaller radii, into regions hot enough to deplete Li, thus, resulting in a lower Li abundance at the age of the Sun. 

The black vertical line in Fig.~\ref{fig:li} denotes the beginning of the MS. Both convection models predict that the Li depletion mainly happens in the pre-MS phase, but not later. Observations of open clusters and solar analogues find Li depletion during the MS phase \citep{Chaboyer1998, Carlos2016, Carlos2020, Mishenina2020, Mishenina2022, Rathsam2023}, which is neither predicted by the \mlt\, nor by the \eq, even if the convective envelope tends to reach deeper than in the \mlt.
Several solutions are suggested in the literature to solve this discrepancy. For example, \citet{Schlattl1999} and \citet{Zhang2012a} studied how to modify the overshooting to bring models and observations in agreement. Furthermore, additional causes of mixing beyond the convectively unstable region, such as gravitational settling \citep{Andrassy2013, Andrassy2015}, internal gravity waves \citep{Montalban1994, Montalban2000, Charbonnel2005} and rotational mixing \citep{Zahn1992, Charbonnel1994, Constantino2021}, are discussed as a cause of the observed Li depletion.
Since these processes are not yet fully understood, and not included in SSMs, the \mlt\ and the \eq\ suffer from the same general shortcoming of SSMs.

\section{General structure of a convective region}
\label{ap:schematic-conv.zone}

\begin{figure}
    \centering
    \includegraphics[width=\linewidth]{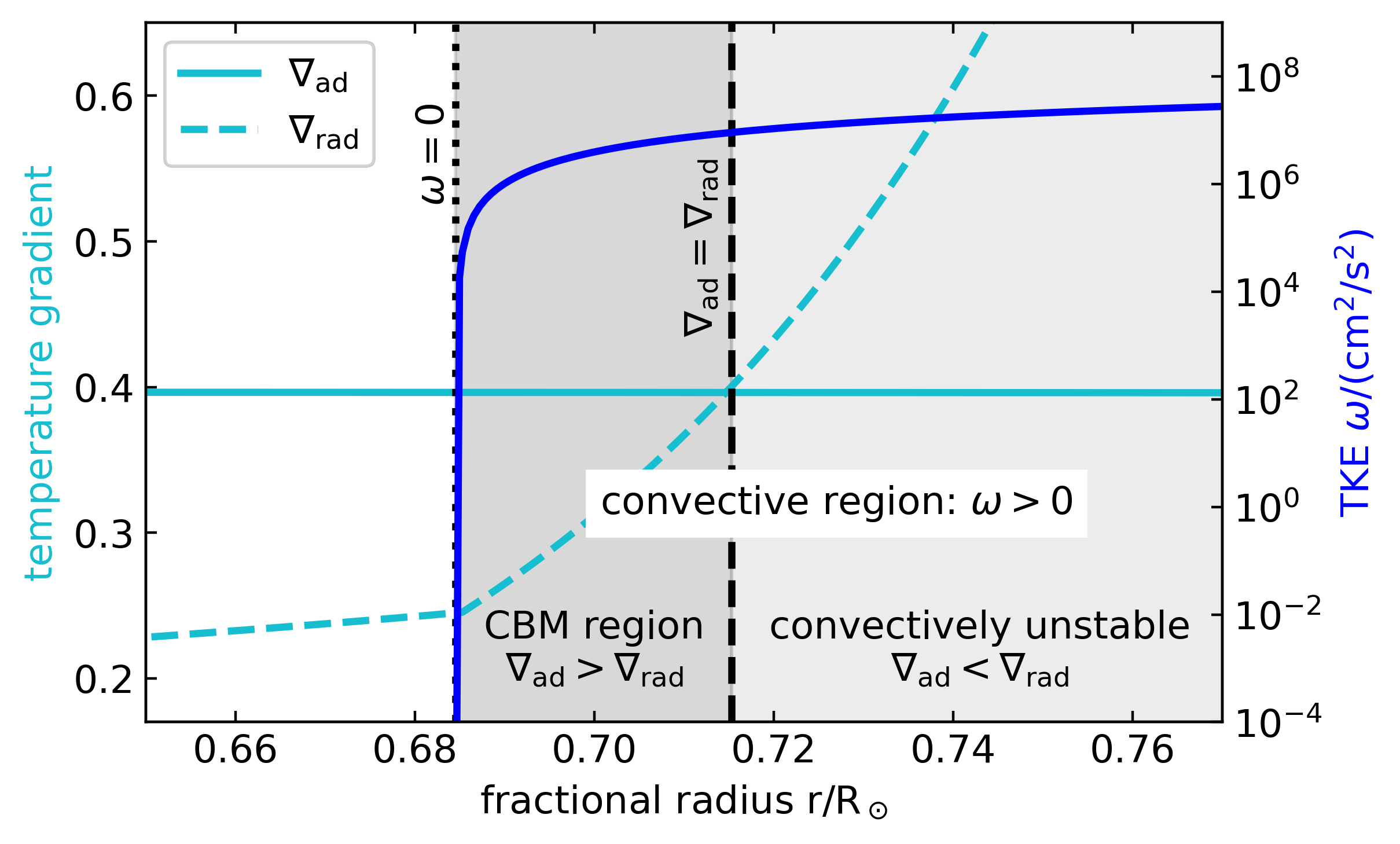}
    \caption{Schematic plot of a convective region. The convective region, where $\omega>0$, is divided into a CBM region and a convectively unstable region. The vertical lines mark the boundary of the convective region (dotted) and the Schwarzschild boundary ($\nabla_\mathrm{ad}=\nabla_\mathrm{rad}$, dashed).}
    \label{fig:schematic-boundaries}
\end{figure}

Figure \ref{fig:schematic-boundaries} shows the structure of a general convective zone, in order to clarify the terminology. When $\omega>0$, there is convective motion, thus, we call this region the convective region. For simple one-dimensional convection models as MLT and the 1-equation model, this region is divided into the convectively unstable region, where $\nabla_\mathrm{ad}<\nabla_\mathrm{rad}$, and the CBM region, where $\nabla_\mathrm{ad}>\nabla_\mathrm{rad}$. The boundary between the convectively unstable region and the CBM region is the Schwarzschild boundary ($\nabla_\mathrm{ad}=\nabla_\mathrm{rad}$). The radius at which the TKE drops to zero is the boundary of the convective region. If a CBM region is present, this is synonymous with the boundary of the CBM region. 
In classical MLT, the TKE drops to zero at the Schwarzschild boundary. Thus, there is no CBM region, and the boundary of the convective region is synonymous with the boundary of the convectively unstable region, which is the Schwarzschild boundary.

%The temperature gradient in the CBM region depends on the description of CBM. 
The temperature gradient depends on the details of the convection theory. 
Ad hoc overshooting only considers the transport of elements but does not include heat transport. Thus, if MLT with ad hoc overshooting is applied, the temperature gradient is not modified compared to the temperature gradient predicted from classical MLT. It is adabatic in the convectively unstable region and becomes radiative at the Schwarzschild boundary. The 1-equation model predicts a marginally subadiabatic temperature gradient in the CBM region, with a sharp transition to a radiative gradient at the boundary of the CBM region (see Fig. \ref{fig:tempgradient}). Other, less simplified convection models and 3D simulations predict a gradual change from an adiabatic to a radiative temperature gradient. The temperature gradient can become subadiabatic already before the Schwarzschild boundary. This gives rise to a Deardorff-layer, which is a region with a subadiabatic temperature gradient and a positive convective flux (see Sect.~\ref{sec:discussion}, and references therein).

\end{appendix}

\end{document}